\documentstyle[prl,aps,epsfig,amssymb]{revtex}
\begin{document}
\draft

\title{Entropic torque}
\author{R. Roth$^{1,2}$, R. van Roij$^3$, D. Andrienko$^4$\thanks{present
address: Max-Planck-Institut f{\"u}r Polymerforschung, Achermannweg 10,
D-55128 Mainz, Germany}, K. R. Mecke$^{1,2}$, and S. Dietrich$^{1,2}$}
\address{$^1$Max-Planck-Institut f{\"u}r Metallforschung, Heisenbergstrasse 1,
D-70569 Stuttgart, Germany}
\address{$^2$Institut f{\"u}r Theoretische und Angewandte Physik, 
Universit{\"a}t Stuttgart, Pfaffenwaldring 57, D-70569 
Stuttgart, Germany} 
\address{$^3$Institute for Theoretical Physics, Utrecht University, 
Princetonpln 5, 3584 CC Utrecht, The Netherlands}
\address{$^4$H.H. Wills Physics Laboratory, University of Bristol, Tyndall 
Avenue, Bristol BS8 1TL, United Kingdom}
\maketitle
\begin{abstract}
Quantitative predictions are presented of a depletion-induced torque and force
acting on a single colloidal hard rod immersed in a solvent of hard spheres 
close to a planar hard wall. This torque and force, which are entirely of 
entropic origin, may play an important role for the key-lock principle, where 
a biological macromolecule (the key) is only functional in a particular 
orientation with respect to a cavity (the lock).
\end{abstract}

The depletion effect, i.e., the effect that smaller particles in a colloidal 
mixture are expelled from those regions where bigger particles are 
sufficiently close together, leads to an effective force between the bigger 
particles. In the case of hard-core interactions these effective forces are 
purely entropic in origin \cite{Asakura54,Vrij76}. For the relatively
simple geometries of two large spheres or a sphere close to a planar or curved
wall in a solvent of small hard spheres this depletion force has been studied 
in detail in theory \cite{Roth99,Goetzelmann99,Roth00}, simulation 
\cite{Biben96,Dickman97}, and experiment \cite{Crocker99,Rudhardt98}. Depletion
forces are crucial for the phase behavior of colloidal system; they can drive 
gas-liquid and liquid-solid phase separation, e.g., in colloid-polymer
\cite{Dijkstra99} and colloid-colloid mixtures \cite{Dijkstra98}. It has been 
suggested \cite{r7} that depletion effects play an important role in the 
biological ``key-lock'' mechanism, in which a nonspherical macromolecule (the
key) fits into a cavity (the lock) and forms a chemical bond, but only if 
the key has a particular orientation. A robust performance of this mechanism
is possible if the biological environment is capable of passively transporting
the key, correctly oriented, to the lock. This requires a chemically unspecific
{\em force} to transport the center-of-mass of the macromolecule towards the
cavity and a {\em torque} to orient it. In order to be able to study the
relevance of the entropic contribution to this mechanism a quantitatively 
reliable tool for predicting such an entropic torque is needed. Using density 
functional theory and computer simulations, we show and predict quantitatively
that depletion effects can generate both a force and a torque on a nonspherical
particle in the vicinity of a planar wall thereby providing further evidence 
for the relevance of depletion phenomena in key-lock mechanisms.

\begin{figure}
\centering\epsfig{file=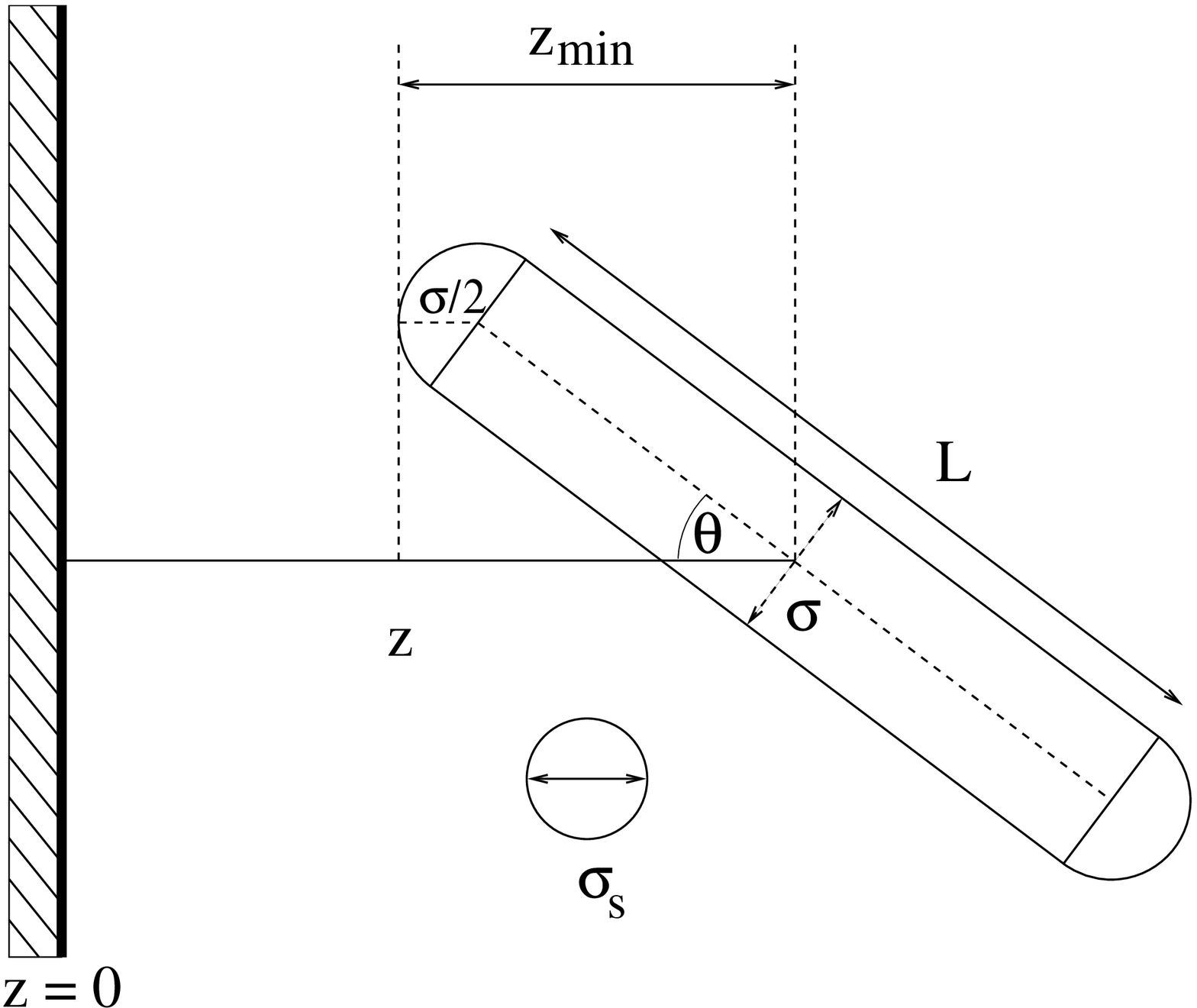,width=0.75\linewidth}
\vspace{0.5cm}
\caption{\label{fig:rod} A spherocylinder of length $L$ and diameter $\sigma$
at angle $\theta$ and distance $z$ relative to a 
a planar hard wall at $z=0$. The minimal value of $z$
is $z_{min}(\theta)=(\sigma+L |\cos\theta|)/2$. The rod is immersed in
a solvent of hard spheres of diameter $\sigma_s$ and number density $\rho_s$ 
far from the wall.}
\end{figure}

The specific system we study is illustrated in Fig.\ref{fig:rod} and consists 
of (i) an infinite planar hard wall located at $z=0$, (ii) a hard 
spherocylinder with cylinder length $L$, diameter $\sigma$, at distance $z$ 
from the wall and forming an angle $\theta \in[0,\pi/2]$ with respect to the 
wall normal, and (iii) a solvent of hard spheres with diameter $\sigma_s$ 
and number density $\rho_s$ in the bulk, i.e., far away from the wall. Since
we are considering only hard interactions the temperature $T$ of the system
plays only the role  of an energy scale via $\beta=1/k_B T$. We are interested
in the effective rod-wall potential $W(z,\theta)$ induced by the spheres. The 
force and the torque on the rod follow by differentiating $-W(z,\theta)$ with 
respect to $z$ and $\theta$, respectively. The bare rod-wall interaction 
restricts the center of mass of the rod to $z$ to $z\geq z_{min}(\theta)$ = 
$(\sigma +L |\cos\theta|)/2$. The system, and hence $\beta W(z,\theta)$, is 
completely characterized by the aspect ratio $L/\sigma$ of the rod, the 
diameter ratio $\sigma/\sigma_s$, and the packing fraction 
$\eta=\rho_s \pi \sigma_s^3/6$.

Following Asakura and Oosawa \cite{Asakura54}, a first approximation
of the effective potential $W(z,\theta)$ can be obtained from the analysis of 
the volume that is excluded to the centers of the solvent spheres by the wall 
and the rod. The wall excludes the slab $0<z<\sigma_s/2$, and the rod the 
volume of a spherocylinder of length $L$ and diameter $\sigma+\sigma_s$. If 
these two contributions to the excluded volume overlap, i.e., for 
$z<z_{min}+\sigma_s$ (see Fig.~\ref{fig:rod}) the volume accessible to the 
centers of the spheres and hence the entropy of the solvent increases. This 
gain of entropy translates directly into an effective, purely attractive force
acting on the center of the rod. An important ingredient of this so-called 
Asakura-Oosawa approximation (AOa) is that the density profile of the spheres 
is taken to be constant and equal to the bulk density $\rho_s$, i.e., all 
correlations in the solvent are entirely neglected. The AOa therefore 
underestimates the strength and the range of the force acting on the rod. 

Here we go beyond the AOa taking the sphere-sphere correlations into account.
We recall that the depletion potential equals the grand-potential change of 
the (inhomogeneous) fluid of spheres in contact with the wall upon moving the 
rod from $z\rightarrow\infty$ to a finite value of $z$, at a given $\theta$. 
Within density functional theory (DFT), with 
${\cal F}_{ex}[\rho_s,\rho_r]$ the excess (over ideal) free energy functional 
of a {\em mixture} of hard spheres and rods, one can write \cite{Roth00}
\begin{equation} \label{pot}
\beta W(z,\theta) = \lim_{\rho_r\to 0} [c_r^{(1)}(z\to\infty,\theta) - 
c_r^{(1)}(z,\theta)],
\end{equation}
where  $c^{(1)}_r(z,\theta)=-\delta \beta {\cal F}_{ex}[\rho_s,\rho_r]/
\delta \rho_r(z,\theta)$ is the direct one-body correlation function of the 
rods \cite{Evans79} where $\rho_r(z,\theta)$ is the number density of rods for
a given orientation $\theta$. The functional we use is based on Rosenfeld's 
fundamental measure theory (FMT) for mixtures of general convex hard bodies 
\cite{Rosenfeld94}. FMT has proved to account accurately for both the structure
and the thermodynamics of inhomogeneous hard-sphere mixtures \cite{Rosenfeld89}
as compared with simulations. Recently, a FMT for a mixture of spheres and 
needles of vanishing thickness was proposed \cite{Schmidt01}. Our extension to
include a nonspherical particle with finite volume (the rod) in the theory 
requires the deconvolution of the Mayer-$f$ function of the rod-sphere 
interaction into a set of orientation dependent weight functions 
\cite{Rosenfeld94}. The details of this technically involved deconvolution 
will be explained elsewhere \cite{Roth02}. 

We obtain the depletion potential in two steps. First we calculate the 
unperturbed density profile $\rho_s(z)$ of the spheres close to the wall. This
is the equilibrium profile in the limit $\rho_r\to 0$, as required in 
Eq.(\ref{pot}). The functional reduces, in this case, to the well-tested FMT 
for spheres. In the second step we use $\rho_s(z)$ to evaluate 
$c_r^{(1)}(z,\theta)$, using the deconvoluted Mayer-$f$ function. From 
Eq.(\ref{pot}) we then obtain the depletion potential $W(z,\theta)$ for all 
$z$ and $\theta$ (from the profile $\rho_s(z)$ alone!).

\begin{figure}
\centering\epsfig{file=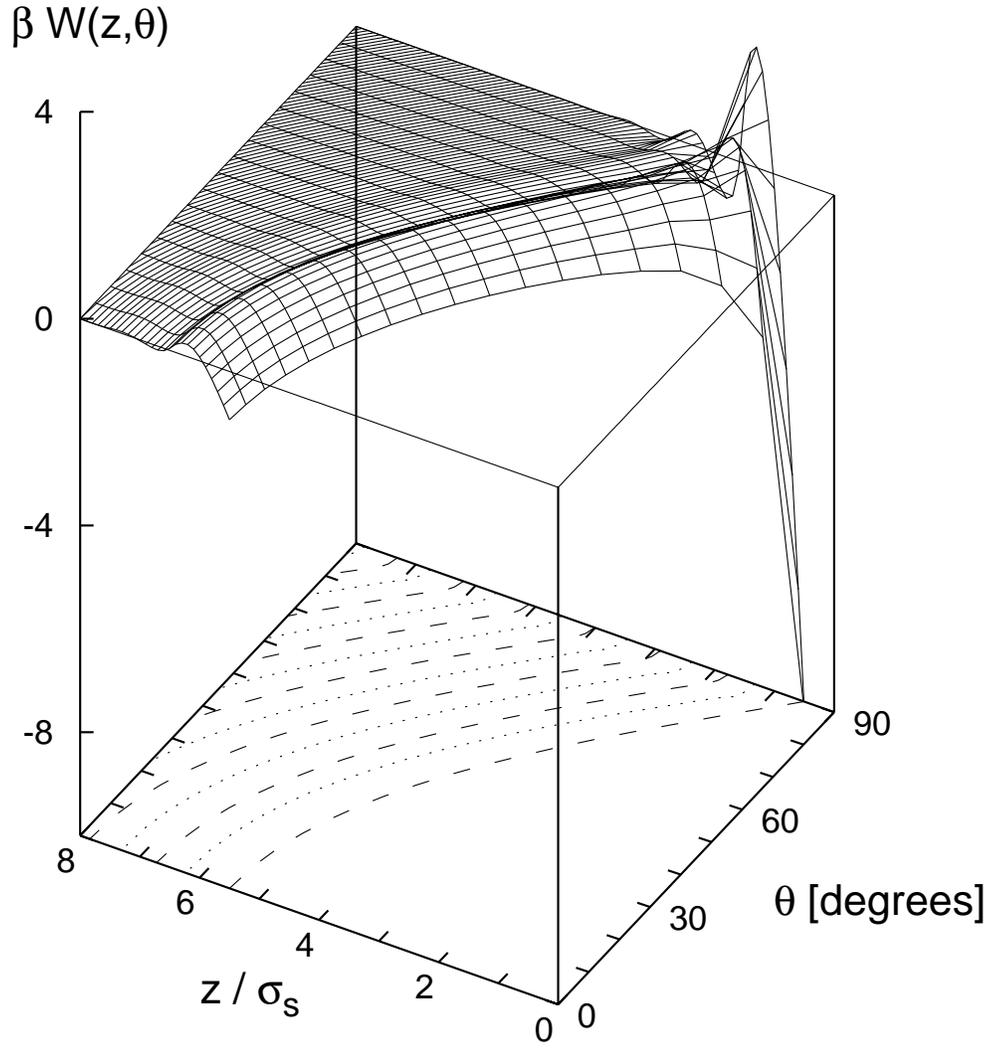,width=0.75\linewidth}
\caption{\label{fig:pot2D} Depletion potential $\beta W(z,\theta)$ for the 
spherocylinder shown in Fig.~\ref{fig:rod} for $L/\sigma=10$, 
$\sigma/\sigma_s=1$, and a packing fraction 
$\eta_s=\rho_s \pi \sigma_s^3/6=0.2239$ of the hard-sphere solvent. $\theta$ 
is measured in degrees. The dashed and dotted lines represent the positions of
local minima and maxima, respectively.}
\end{figure}

We present results for $L/\sigma=10$, $\sigma/\sigma_s=1$, and a packing 
fraction $\eta_s=(\pi/6)\sigma_s^3\rho_s=0.2239$ for the spheres. 
Figure~\ref{fig:pot2D} displays the depletion potential $\beta W(z,\theta)$ as
function of $z$ and $\theta$. The first observation is that the potential is, 
unlike the AOa, {\em not} monotonic; the hard-sphere correlations generate a 
sequence of potential barriers and wells. The dashed and dotted curves in 
Fig.~\ref{fig:pot2D} denote the positions of the minima and maxima of the 
potential. We note that for small angles $\theta\ll 1$ the shape of the 
rod-wall depletion potential $W(z,\theta)$ coincides almost perfectly with the
depletion potential $W_{ws}(z)$ between a sphere of diameter $\sigma$ and the 
wall, i.e., $W(z,\theta\ll 1)\approx W_{ws}(z-L \cos|\theta|/2)$. For these 
small angles the length of the rod $L$ is rather unimportant for details of 
the depletion potential, as we have verified for various values of $L/\sigma$ 
and $\sigma/\sigma_s$. For large angles, however, the whole geometry of the 
rod, i.e., $L/\sigma$ and $\sigma/\sigma_s$, is relevant, and the depth of the 
depletion potential at contact becomes more negative as both $L/\sigma$ and 
$\sigma/\sigma_s$ are increased. In addition the contact value of the depletion
potential can be further decreased by increasing $\eta_s$.

\begin{figure}
\centering\epsfig{file=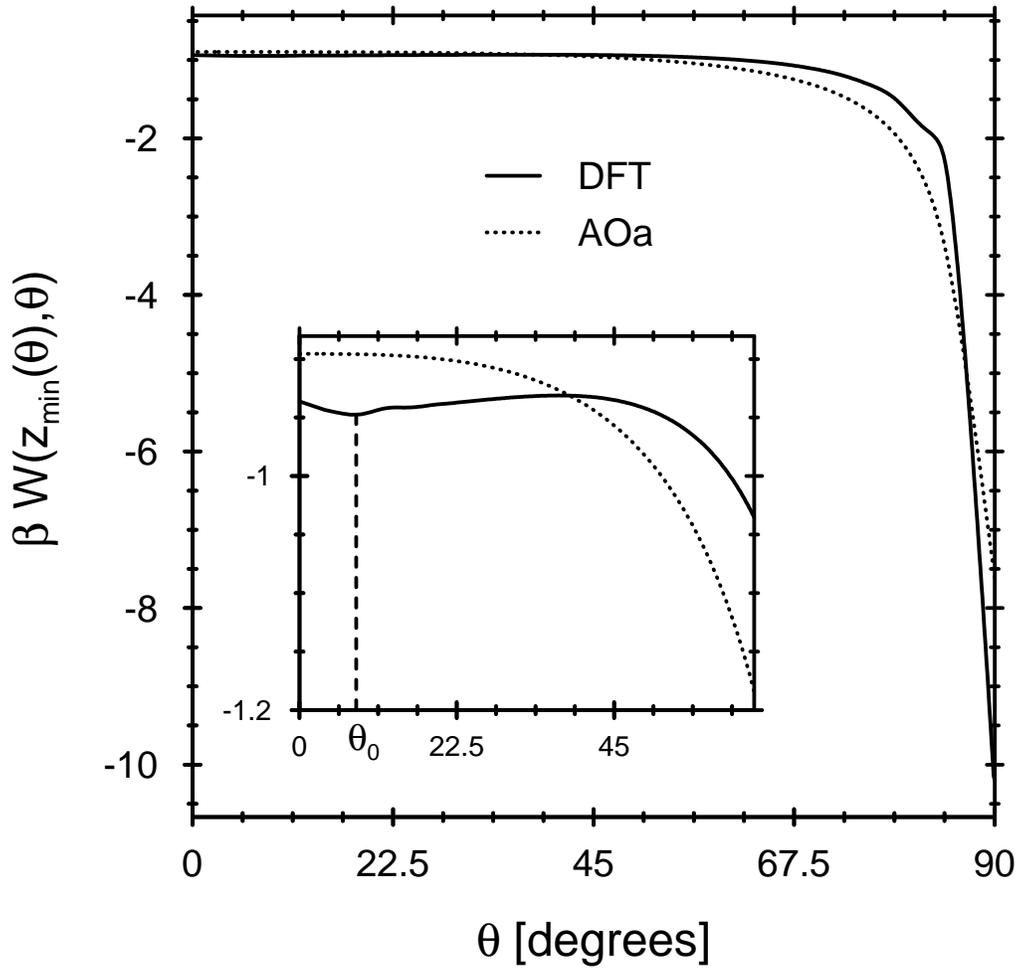,width=0.75\linewidth}
\vspace{0.5cm}
\caption{\label{fig:w0} The contact value $\beta W(z_{min}(\theta),\theta)$ as
a function of $\theta$ (in degrees) as calculated within DFT (full line) and 
within the AOa (dotted line) for the system described in Fig.~\ref{fig:pot2D}. 
The inset shows that a rod in contact with the wall can exhibit a metastable
orientation for $\theta=\theta_0\approx 8$ degrees with a barrier height 
$\beta \Delta W=0.007$.}
\end{figure}

Although certain general trends of the influence of the geometry on the shape 
of the depletion potential can already be roughly understood within the simple
AOa \cite{Asakura54}, the correlations in the hard-sphere fluid are very 
important and can lead to a quantitatively and even qualitatively different 
behavior. In Fig.~\ref{fig:w0} we illustrate this by comparing the depletion
potential at contact as obtained within our DFT approach
($W(z_{min}(\theta),\theta)$, full line) with the corresponding quantity within
the AOa ($W^{AOa}(z_{min}(\theta),\theta)$, dotted line). While for small 
and large angles the contact value of the full depletion potential is more 
negative than within the AOa, for intermediate values of $\theta$ packing 
effects can shift the actual contact value above its corresponding AOa-value. 
If the rod is almost parallel to the wall, i.e., for $\theta \lesssim \pi/2$ 
the deviation of the contact value calculated within the AOa from the 
prediction of the full DFT is significant and much more pronounced than in the 
well-studied wall-sphere or sphere-sphere geometry. The inset of 
Fig.~\ref{fig:w0} shows that a rod in contact with the wall can exhibit a
metastable orientation for a certain angle $0<\theta_0<\pi/2$ while the AOa 
predicts that this configuration is unstable. Although for the parameters 
under consideration here the height of the potential barrier is rather small 
($0.007 k_B T$), we note that its value increases for smaller values of 
$\sigma/\sigma_s$.

As a consequence of the dependence of the rod-wall depletion potential on the 
orientation $\theta$ for a given (fixed) distance $z$ of the center of the rod
from the wall an entropic torque acts on the rod which drives it into an 
orientation with minimal depletion potential. This torque can be estimated
by replacing the spherocylinder by a dumbbell composed of two spheres with
diameter $\sigma$ connected by an infinitesimally thin but rigid wire of length
$L$. Since the sphere-wall depletion force $f_{sw}(z)$ acts on the two spheres
located at $z_\pm = z \pm (L/2) \cos \theta$ the corresponding torque with
respect to the center of mass is given by
\begin{equation} \label{simpletorque}
M_{db}(z,\theta) = \frac{L}{2} \sin\theta \left[ f_{sw}(z_-) -
f_{sw}(z_+) \right].
\end{equation}
It turns out that at low packing fractions of the hard-sphere solvent, or at 
sufficiently large separations of the rod from the wall 
Eq.~(\ref{simpletorque}) yields a semi-quantitative expression for the torque 
acting on a spherocylinder.

\begin{figure}
\centering\epsfig{file=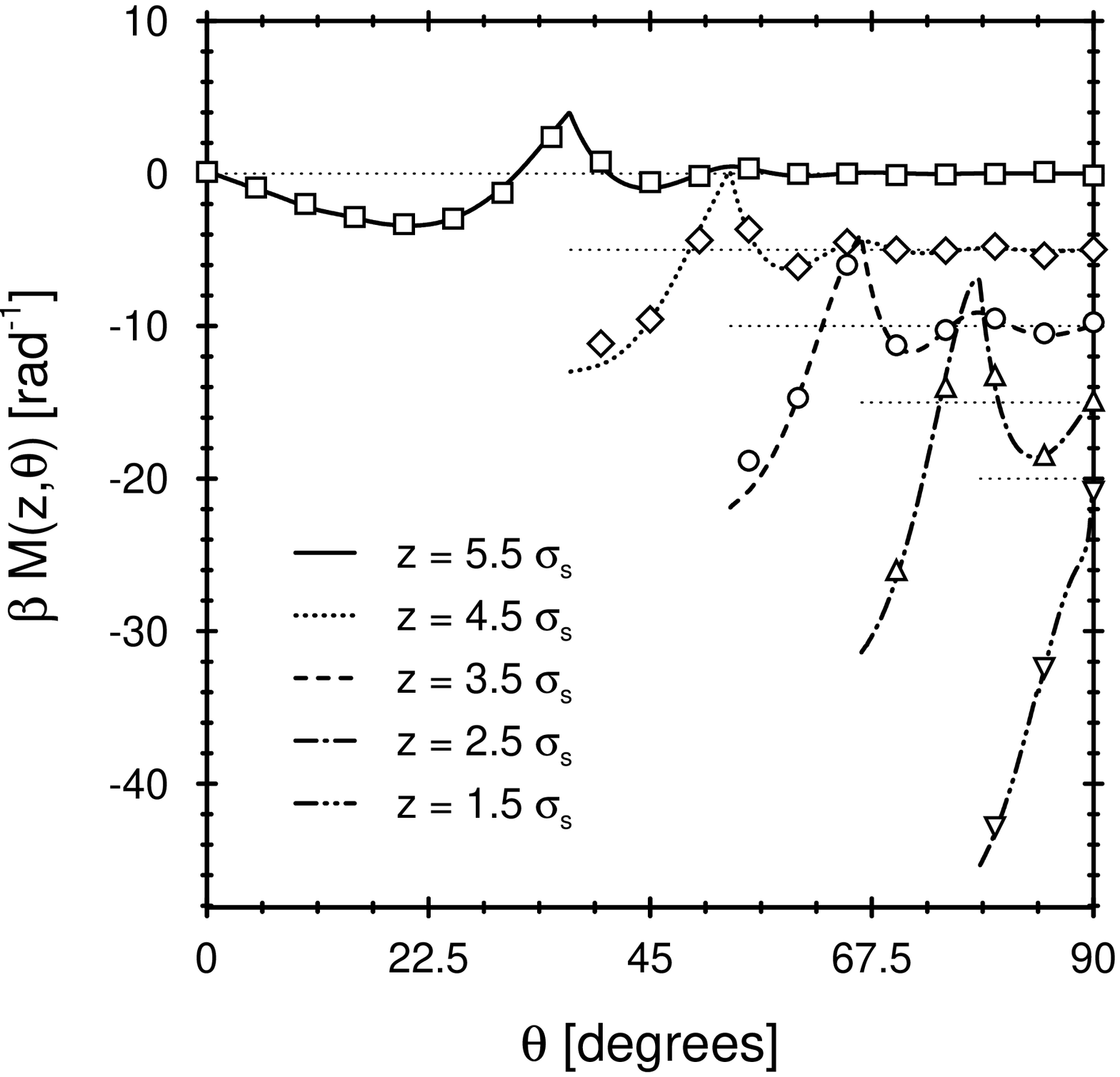,width=0.75\linewidth}
\caption{\label{fig:torque} The torque $\beta M(z,\theta)$ as function of 
$\theta$ for various values of $z$. For reasons of clarity the curves for
$z=4.5$, $3.4$, $\dots$ are shifted downward by $-5$, $-10$, $\dots$,
respectively. The horizontal lines $M=0$ are indicated partially. The 
parameters of the system are the same as in Fig.~\ref{fig:pot2D}. Because of 
the hard wall at $z=0$ the torque is defined only for 
$\theta \geq \arccos(2 z/L-\sigma/L)$ if $z<(\sigma+L)/2$. $M>0$ $(<0)$ 
corresponds to a force which tends to align the rod parallel (normal) to 
the wall. The lines denote our DFT results which are in excellent agreement 
with our simulations (symbols). The error bars of the simulations are of the 
order of the symbol size.}
\end{figure}

From the full rod-wall depletion potential $W(z,\theta)$ one can obtain the 
torque by rotating the rod by an infinitesimal angle $d\theta$ around an axis 
through the center of the rod in a direction characterized by the unit vector 
${\bf n}_\theta$ normal to the symmetry plane shown in Fig.~\ref{fig:rod}. The
corresponding change in the depletion potential is $d W(z,\theta)$ which can be
written as $d W=- \sum_i {\bf f}_i\cdot d{\bf r}_i$, i.e., as a sum of forces 
${\bf f}_i$ acting on the rod at positions ${\bf r}_i$ from the center with 
$d {\bf r}_i=d \theta ({\bf r}_i\times {\bf n}_\theta$). It follows that the
torque ${\bf M}(z,\theta)=M(z,\theta) {\bf n}_\theta$ with respect to the 
center of mass is given by
\begin{equation}
M(z,\theta) = - \frac{\partial W(z,\theta)}{\partial \theta}.
\end{equation}
The symmetry of the problem leads to $M(z,\theta)=0$ for $\theta=0$ and
$\theta=\pi/2$. A positive value of the torque acts on the rod as to
increase the angle $\theta$ (rotating it parallel to the wall) while a 
negative value of $M$ leads to a decrease of $\theta$ (rotating it normal to
the wall). Some typical examples for the torque as a function of $\theta$ for 
various values of $z$ are shown in Fig.~\ref{fig:torque}. Lines and symbols 
denote results from DFT and molecular dynamics (MD) simulations, respectively.
The DFT predictions are in excellent agreement with the simulations. The 
details of the simulations will be presented elsewhere \cite{Roth02}.

For larger distances of the rod from the wall $W(z,\theta)$ exhibits local 
minima and maxima in addition to the global minimum with corresponding zeros 
of the torque (see Fig.~\ref{fig:torque} for $z=5.5 \sigma_s$ and 
$z=4.5 \sigma_s$). A minimum (maximum) of $W(z,\theta)$ leads to a zero of 
$M(z,\theta)$ with a negative (positive) gradient in $\theta$. Furthermore, the
torque can exhibit a cusp (see Fig.~\ref{fig:torque}) if the minimal distance
between the wall and the rod leaves space for precisely one solvent sphere.
This can be realized for distances 
$\sigma_s+\sigma/2 \leq z \leq  \sigma_s + (L+\sigma)/2$ and a corresponding
orientation $\theta_{cusp}=\arccos((2 (z-\sigma_s)-\sigma)/L)$.

As the rod moves closer to the wall the modulus of the torque increases and 
for small separations from the wall (see, e.g., $z=1.5 \sigma_s$ in 
Fig.~\ref{fig:torque}) the torque vanishes only for $\theta=\pi/2$ which is, 
however, only a metastable configuration. For such small and fixed values of 
$z$ the entropic torque rotates the rod towards configurations with smaller 
angles until the rod is in contact with the wall and has to stop its rotation.
Thus the rod reaches its most favorable configuration of lying parallel in
contact with the wall not by approaching the wall in a parallel configuration
but by touching the wall first at one end and then by decreasing the distance 
of its center from the wall. The modulus of the maximum of the entropic torque
for the system considered here is of the order of about $20 k_B T$ rad$^{-1}$, 
which corresponds to roughly $10^{-20}$ J rad$^{-1}$ at room temperature. The 
strength of the torque increases for larger values of $L/\sigma$, 
$\sigma/\sigma_s$, or $\eta_s$. For example, for $L/\sigma=20$, 
$\sigma/\sigma_s=2$, and $\eta_s=0.2239$, the modulus of the maximum of the 
entropic torque reaches a value of about $38.7 k_B T$ rad$^{-1}$. We note that
the maximum torque acts at small values of $z$ for which the AOa also would 
predict the existence of a torque; however, its magnitude is largely enhanced 
through the correlations in the solvent.

With the results presented so far we can comment on some aspects of the path 
in the $z$-$\theta$ plane a rod would take upon approaching the wall from the 
bulk. Only if the rod comes sufficiently close to the wall to be subject to the
oscillations of the number density of the hard-sphere fluid as a function of 
$z$, the entropic force and torque will act on it. The closer the rod gets to 
the wall the higher are the potential barriers between the minima (see 
Fig.~\ref{fig:pot2D}) the rod has to overcome by thermal motion in order to 
move still closer. The potential barriers are, however, only moderate and easy
to overcome for small angles and increase for angles $\theta \gtrsim 60$ 
degrees so that it is not optimal to rotate the rod until it is parallel to the
wall since then the potential barrier for further approach is  largest and for
the parameters considered here more than $2.3 k_B T$. On the other hand along
the line of the second minimum -- see the dashed lines in Fig.~\ref{fig:pot2D}
-- $\beta W$ has a similar shape as $\beta W(z_{min}(\theta),\theta)$ [see
Fig.~\ref{fig:w0}], so that the entropic torque will rotate the rod into the 
local minimum at $\theta=\pi/2$ if $\theta$ is already sufficiently large. Once
the rod is in contact with the wall the torque will rotate it towards the 
pronounced global minimum of $\beta W(z,\theta)$.

We have presented quantitative predictions of the entropic torque acting on a
hard spherocylinder close to a hard planar wall in a solvent of small hard 
spheres. Our DFT predictions are in excellent agreement with our MD 
simulations. We find that the the depletion effect leads to a significant
entropic torque and it is tempting to speculate that this entropic torque can
play an important role in understanding the key-lock principle.

It is a pleasure to thank R. Evans, M. Schmidt, C. Bechinger, and M. P. Allen 
for stimulating discussions. This work is part of the Research program of the 
``Stichting voor Fundamenteel Onderzoek der Materie (FOM)'', which is 
financially supported by the ``Nederlandse Organisatie voor Wetenschappelijk 
Onderzoek (NWO)''. D.A. acknowledges the support of the Overseas Research
Students Grant, EPSRC grants GR/L89990, GR/M16023, INTAS grant 99-00312.
MD simulations used the GBMEGA program of the `Complex Fluids Consortium' with
computer time allocated at the CSAR facility.

\newpage

\begin{references}
\bibitem{Asakura54} S. Asakura and F. Oosawa, J. Chem. Phys. {\bf 22}, 1255 
(1954); J. Polymer Sci. {\bf 33}, 183 (1958).
\bibitem{Vrij76} A. Vrij, Pure and Appl. Chem. {\bf 48}, 471 (1976).
\bibitem{Roth99} R. Roth, B. G\"otzelmann, and S. Dietrich, Phys. Rev. Lett.
{\bf 83}, 448 (1999).
\bibitem{Goetzelmann99} B. G\"otzelmann, R. Roth, S. Dietrich, M. Dijkstra, 
and R. Evans, Europhys. Lett. {\bf 47}, 398 (1999); R. Roth, doctoral thesis, 
Bergische Universit\"at Wuppertal (1999).
\bibitem{Roth00} R. Roth, R. Evans, and S. Dietrich, Phys. Rev. E {\bf 62},
5360 (2000).
\bibitem{Biben96} T. Biben, P. Bladon, and D. Frenkel, J. Phys.: Condens. 
Matter {\bf 8}, 10799 (1996).
\bibitem{Dickman97}R. Dickman, P. Attard, and V. Simonian, J. Chem. Phys. 
{\bf 107}, 205 (1997).  
\bibitem{Crocker99} J.C. Crocker, J.A. Matteo, A.D. Dinsmore, and 
A.G. Yodh, Phys. Rev. Lett. {\bf 82}, 4352 (1999).
\bibitem{Rudhardt98} D. Rudhardt, C. Bechinger, and P. Leiderer, Phys. Rev. 
Lett. {\bf 81}, 1330 (1998).
\bibitem{Dijkstra99} M. Dijkstra, J.M. Brader, and R. Evans, J. Phys.: 
Condens. Matter {\bf 11}, 10079 (1999).
\bibitem{Dijkstra98} M. Dijkstra, R. van Roij, and R. Evans, Phys. Rev. Lett. 
{\bf 81}, 2268 (1998); {\em ibid} {\bf 82}, 117 (1999); Phys. Rev. E {\bf 59}, 
5744 (1999).
\bibitem{r7} see, e.g., M. Kinoshita and T. Oguni, Chem. Phys. Lett. {\bf 351},
79 (2002) and references therein.
\bibitem{Evans79} R. Evans, Adv. Phys. {\bf 28}, 143 (1979).
\bibitem{Rosenfeld94} Y. Rosenfeld, Phys. Rev. E {\bf 50}, R3318 (1994);
Mol. Phys. {\bf 86}, 637 (1995).
\bibitem{Rosenfeld89} Y. Rosenfeld, Phys. Rev. Lett. {\bf 63}, 980 (1989);
J. Chem. Phys. {\bf 98}, 8126 (1993).
\bibitem{Schmidt01} M. Schmidt, Phys. Rev. E {\bf 63}, 050201-1 (2001).
\bibitem{Roth02} R. Roth, R. van Roij, D. Andrienko, K. R. Mecke, and S. 
Dietrich, unpublished.
\end{references}
\end{document}